\documentclass{appolb}
\usepackage{graphicx} % support the \includegraphics command and options
\usepackage{amsmath} %maths symbols
\usepackage{amssymb} %maths symbols
\usepackage{bbm} %maths symbols
\usepackage{amsthm}
\usepackage{authblk}
\numberwithin{equation}{section}
\newtheorem{thm}{Theorem}
\newtheorem{lemma}{Lemma}

\newcommand{\bea}{\begin{eqnarray}}
\newcommand{\eea}{\end{eqnarray}}

\newcommand{\Lv}{\boldsymbol L}
\newcommand{\Rv}{\boldsymbol R}
\newcommand{\C}{\mathbb{C}}
\newcommand{\E}{\mathbb{E}}

\newcommand{\tr}{\mbox{Tr}}

\newcommand{\xb}{\bar{x}}

\newcommand{\lambdab}{\bar{\lambda}}

\begin{document}
\title{Determinantal structure and bulk universality of conditional overlaps in the complex Ginibre ensemble%
\thanks{Presented by the first author at ``Random Matrix Theory: Applications in the era of information theory'', April 29 - May 3 2019, 
Krak\'ow, Poland}%
}
\author{Gernot Akemann
\address{
Faculty of Physics,
Bielefeld University,
P.O. Box 100131,
33501 Bielefeld, Germany,\\
Department of Mathematics, Royal Institute of Technology (KTH),\newline Brinellv\"agen 8, 114 28 Stockholm, Sweden,}
\\
{Roger Tribe, Athanasios Tsareas, and Oleg Zaboronski,}
\address{Department of Mathematics, University of Warwick, CV4~7AL
Coventry, UK.}
}

\maketitle
\begin{abstract}
In these proceedings we summarise how the determinantal structure for the conditional overlaps among left and right eigenvectors emerges in the complex Ginibre ensemble at finite matrix size. 
An emphasis is put on the underlying structure of orthogonal polynomials in the complex plane and its analogy to the determinantal structure of $k$-point complex eigenvalue correlation functions.
The off-diagonal overlap is shown to follow from the diagonal overlap conditioned on $k\geq2$ complex eigenvalues. As a new result we present the local bulk scaling limit of the conditional overlaps away from the origin. It is shown to agree with the limit at the origin and is thus universal within this ensemble. 
\end{abstract}

%%%%%%%%%%%%%%%%%%%%%%%%%%%%%%%%%%%%%%%%%%%%%%%%%%  
\section{Introduction}

The motivation to study the statistics of eigenvectors of random matrices comes from many different directions. In \cite{chalker1} Chalker and Mehlig presented two important features of eigenvectors of non-Hermitian operators, 
characterised by the overlap between left and right eigenvectors:
their r\^ole  in the extreme sensitivity of the complex spectrum of such operators and in transient behaviour in the time evolution of complex dynamical systems. The former has developed into a branch of mathematics under the title of pseudospectra, cf. \cite{TrefethenEmbree}, whereas the latter has been advocated, e.g. in the modelling of random neural networks \cite{Evaetal}. More traditional applications in physics include the line width of lasers 
in a chaotic cavity \cite{Carlo}  as well as scattering in microwave cavities \cite{YanDima}, that have been measured in \cite{Gros}.

One of the salient features of Random Matrix Theory (RMT) is its underlying integrable structure. It has enabled an analytic study of many aspects  of spectral correlations, which are relevant in a large number of applications, cf. \cite{GMW,ABF}. One of our main motivations was to find out whether or not such an integrable structure also exists when considering eigenvectors, possibly at finite matrix size $N$. 
A multitude of techniques has been developed to tackle questions about eigenvalues, and so it is not surprising that these have been also employed in the context of eigenvectors. What is perhaps surprising is that only rather recently have we seen much progress in attacking the questions posed by Chalker and Mehlig \cite{chalker1}. 

Below we give an incomplete list of results for different ensembles of random matrices, with the complex Ginibre ensembles introduced in \cite{ginibre} being most studied. 
Using a combination of Green's functions and diffusion equations, it  was noticed early on that the Dysonian dynamics in this ensemble couples the complex eigenvalues and their eigenvectors in a non-trivial way \cite{BurdaPRL,Burda+}. These techniques were further developed including Feynman diagrams \cite{BSV,MAN-WT}, free probability \cite{Speicher} or stochastic differential equations \cite{GW} and applied to different  ensembles including products of elliptic Ginibre matrices \cite{BSV}. These, as well as truncated unitary and spherical ensembles, were analysed in \cite{GD2} using probabilistic means, after an earlier breakthrough for these methods in \cite{Bourgade}, see also \cite{FlorentOfer} for the correlations between angles of eigenvectors.  The quaternionic Ginibre ensemble appeared more recently from a probabilistic angle \cite{BGC,GD} as well as for finite-$N$ in \cite{AFK}, using the heuristic tools of \cite{chalker2}.  An entirely different approach uses supersymmetry \cite{Yan} or orthogonal polynomials \cite{FGS}, expressing the relevant quantities in terms of expectation values of characteristic polynomials. This includes also eigenvectors  of real eigenvalues of the real Ginibre ensemble \cite{Yan,YVF-WT}. 

A common underlying question is that of universality of the newly found eigenvectors correlations. While much of this remains open, numerical checks \cite{Bourgade} strongly suggest some universality, and we refer to \cite{MAN-WT} for a comprehensive list of various ensembles in the global bulk regime, pointing at parallels and differences.
A further indication is the recently found universality of complex bulk and edge {\it eigenvalue} correlations away from the real line, uniting all three Ginibre ensembles \cite{BorodinSinclair,AKMP}.
The present work is based on \cite{ATTZ} where we use the technique of orthogonal polynomials in the complex plane, combined with moment methods developed earlier in \cite{WS,Crawford}. 

The following sections are 
organised as follows. In Section \ref{GinUE} we recall relevant features of the complex Ginibre ensemble, including the definition of complex eigenvalue and overlap correlation functions. Section \ref{finiteN} summarises our discovery of an integrable structure at finite-$N$, giving determinantal formulae for the conditional diagonal and off-diagonal overlaps. This exploits an exact relation between the two.  
For the off-diagonal overlap more details are given in \cite{ATTZ}. In Section \ref{bulklargeN} we focus on the local statistics of the diagonal overlap everywhere in the bulk of the spectrum, extending the results for the origin from \cite{ATTZ}. For further results regarding edge statistics, the limiting connection between edge and bulk as well as for large argument separation in the bulk we refer also to \cite{ATTZ}. Our conclusion and discussion of open problems is presented in Section \ref{disc}.

%%%%%%%%%%%%%%%%%%%%%%%%%%%%%%%%%%%%%%%%%%%%%%%%%%  
\section{The complex Ginibre ensemble and definition of conditional overlaps}\label{GinUE}

Let us recall the definition of the complex Ginibre ensemble. It consists of 
matrices $M$
of size $N\times N$ with its independent complex Gaussian entries distributed according to 
\bea
P(M)=\pi^{-N^2}\exp\left[- \tr MM^\dag\right]\ ,
\label{GinDef}
\eea
where $\dagger$ stands for Hermitean conjugation.
The left $\Lv_{\alpha}$ and right eigenvectors $\Rv_{\alpha}$ with complex eigenvalues $\lambda_{\alpha}, \ 1\leq \alpha\leq N$, are defined by 
\bea
\Lv^\dagger_{\alpha} M&=&\lambda_{\alpha} \Lv^\dagger_{\alpha},
\nonumber\\
M\Rv_{\alpha} &=&\lambda_{\alpha} \Rv_{\alpha},~1\leq \alpha \leq N\ .
\eea
They form a bi-orthogonal set with respect to the Hermitean inner product $\langle\cdot,\cdot\rangle$ on $\C^N$:
\bea
\langle \Lv_{\alpha}, \Rv_{\beta} \rangle=\delta_{\alpha,\beta},~1\leq \alpha,\beta \leq N.
\eea
However, left and right eigenvectors are not orthogonal any more
\bea
\langle \Lv_\alpha, \Lv_\beta \rangle\neq 0 \neq\langle\Rv_\alpha, \Rv_\beta\rangle ,~1\leq \alpha<\beta \leq N\ ,
\eea
in contrast to Hermitian RMT. Following \cite{chalker1}, the matrix of overlaps between left and right eigenvectors is then defined as
\bea\label{ovmat}
O_{\alpha \beta}=\langle \Lv_{\alpha}, \Lv_{\beta} \rangle 
\langle \Rv_{\alpha}, \Rv_{\beta} \rangle,~1\leq \alpha,\beta \leq N\ ,
\eea
where the choice of this combination is motivated by its invariance under a simultaneous rescaling of the eigenvectors $\forall\alpha$: $\Rv_{\alpha}\to c\Rv_{\alpha},\  \Lv_{\alpha}\to \Lv_{\alpha}/c$, for any complex $c\neq0$.

The joint density of eigenvalues $p_N(\Lambda)$ of $M$ 
can be found via a Schur decomposition, $M=U(\Lambda+T)U^\dag$, with $U\in U(N)/U(1)^N$ unitary, $T$ complex strictly upper triangular, and $\Lambda=\mbox{diag}(\lambda_1,\ldots,\lambda_N)$ containing the complex eigenvalues:
\bea\label{cginibrelaw}
p_N\left(\Lambda\right)=\frac{1}{Z_N} \left|\Delta^{(N)}(\lambda_1,\ldots,\lambda_N)\right|^2e^{-\sum_{j=1}^N |\lambda_j|^2}.
\eea
Here, $\Delta^{(N)} (\lambda_1,\ldots,\lambda_N)=\prod_{i>j}^N(\lambda_i-\lambda_j)$ is the Vandermonde determinant of $N$ variables,  and the normalising partition function is given by 
\bea
Z_N&=&\int\limits_{\C^{N}} \prod_{i=1}^N d\lambda_i d\lambdab_i\,
|\Delta^{(N)}(\lambda_1,\ldots,\lambda_N)|^2e^{-\sum_{j=1}^N |\lambda_j|^2}
= \pi^N\prod_{j=0}^N j! \quad
\label{ZN}
\eea
Eq. \eqref{cginibrelaw} constitutes a determinantal point process. Recalling the definition of the $k$-point eigenvalue $(ev)$ correlation function, it holds that
\bea\label{cgin1}
\rho^{(N,k)}(\lambda_1,\ldots,\lambda_k)&=&\frac{N!}{(N-k)!} \int\limits_{\C^{N-k}} \prod_{i=k+1}^N d\lambda_i d\lambdab_i\  p_{N}(\Lambda)
\\
&=&
\det_{1\leq i,j\leq k}
\left[K^{(N)}_{ev}(\lambda_i,\lambda_j)
\right],
\nonumber
\eea
where the kernel of orthogonal polynomials reads at finite-$N$
\bea\label{cgin2}
K^{(N)}_{ev}(x,y)=e^{-|x|^2}\sum_{m=0}^{N-1}\frac{(\bar{x}y)^m}{\pi m!},
\eea
see \cite{ginibre,mehta} for details. It is often written in a more symmetric fashion, using the invariance $K_{ev}^{(N)}(x,y)\to (f(x)/f(y))K_{ev}^{(N)}(x,y)$ of the determinant in \eqref{cgin1}, when choosing $f(x)=e^{+|x|^2/2}$. The corresponding monic orthogonal polynomials of the rotationally invariant Gaussian weight $e^{-|z|^2}$ are the monomials $z^k$, $k=0,1,2,\ldots$, with (squared) norms $h_k$,
\bea
\int_{\C}dzd\bar{z}\ z^k\bar{z}^j e^{-|z|^2}= \delta_{j,k} h_j\ ,\quad \mbox{with} \quad h_j=\pi j!
\label{GinOP}
\eea 
The Andr\'ei\'ef integral formula valid for integrable functions $\phi_i(x)$ and $\psi_i(x)$, for $i=1,\ldots,N$,
\bea
\prod_{i=1}^N\int_{\C}\!  dz_id\bar{z}_i\!\det_{1\leq k,l\leq N}[\phi_k(z_l)]\!\det_{1\leq k,l\leq N}[\psi_k(\bar{z}_l)]\!
&=&\! N!\! \det_{1\leq k,l\leq N}\!\left[ \int_{\C}\!dzd\bar{z} \phi_k(z)\psi_{l}(\bar{z})\right] 
\nonumber\\
\label{Andreief}
\eea
then immediately leads to the normalisation 
$Z_N=N! \prod_{j=0}^{N-1}h_j$, as previously stated in \eqref{ZN}.

One of the main results of \cite{chalker2} is that the conditional diagonal $D_{11}^{(N,1)}(\lambda)$ 
can be expressed as an expectation value with respect to the joint density $p_N(\Lambda)$ \eqref{cginibrelaw} alone, after integrating out the upper triangular matrix $T$ in a recursive manner \cite{chalker2}:
\bea
D_{11}^{(N,1)}(\lambda)&=&\E_N\left(\sum_{\alpha=1}^N O_{\alpha \alpha}\delta(\lambda_\alpha-\lambda)\right)
\label{diagO-def}
\\
&=& \int\limits_{\C^{N}}\prod_{i=1}^{N}\!d\lambda_i
d\bar{\lambda}_i\, p_N(\Lambda)
\sum_{\alpha=1}^N\delta(\lambda_\alpha-\lambda)
\prod_{\ell\neq \alpha}^N \!\left[1+\frac{1}{|\lambda_\alpha-\lambda_\ell|^2}\right],
\nonumber
\eea
where $\E_N$ is the expectation with respect to \eqref{GinDef} on the level of matrices. 
For the off-diagonal overlaps 
\bea
D_{12}^{(N,2)}(\lambda,\mu)&=&\E_N\left(\sum_{\alpha\neq \beta=1}^N O_{\alpha \beta}\delta(\lambda_\alpha-\lambda)\delta(\lambda_\beta-\mu)\right),
\label{offdiagO-def}
\eea
a similar expression holds, see \eqref{eqnd12int} below.
The same mechanism applies to the overlaps in the  quaternionic Ginibre ensemble \cite{GD,AFK}. In the real ensemble \cite{Yan} the Laplace transformed joint density of overlap and conditional eigenvalue is given by an averaged ratio of characteristic polynomials, thus only depending on the eigenvalues too.

Using this results of \cite{chalker2}, in analogy to the $k$-point correlation functions \eqref{cgin1}, we introduce the $k$-th diagonal overlap $D_{11}^{(N,k)}$ conditioned on $k\geq1$ eigenvalues, compared to \eqref{diagO-def} for $k=1$:\footnote{In slight abuse of notation we omit that the $D_{11}^{(N,k)}$ also depend on the complex conjugated variables $\bar{\lambda}_1,\ldots,\bar{\lambda}_k$, as the $k$-point functions  \eqref{cgin1} do.}
\bea
D_{11}^{(N,k)}(\lambda_1,\ldots,\lambda_k)&=&\frac{N!}{(N-k)!}\int\limits_{\C^{N-k}}\prod_{i=k+1}^{N}\!d\lambda_i
d\bar{\lambda}_i p_N(\Lambda)
\prod_{\ell=2}^N \!\left[1+\frac{1}{|\lambda_1-\lambda_\ell|^2}\right]
\nonumber\\
&=&\frac{e^{-|\lambda_1|^2}N!}{Z_N(N-k)!}\int\limits_{\C^{N-k}}\prod_{i=k+1}^{N}\!
d\lambda_i d\bar{\lambda}_i
|\Delta^{(N-1)}(\lambda_2,\ldots, \lambda_N)|^2 \nonumber\\
&&\qquad\qquad\qquad\qquad\times \prod_{m=2}^N \pi \omega( \lambda_m,\bar{\lambda}_m\mid \lambda_1,\bar{\lambda}_1).
\label{eqnd11int}
\eea
It agrees with \eqref{diagO-def} for $k=1$, after using the symmetry of $p_N(\Lambda)$ under permutation of indices.
Here, we have also  introduced a new weight function on $\C^3$:
\bea\label{eqnwt}
\omega(z,x|u,v)=\frac{1}{\pi}\left(1+(z-u)(x-v)\right)e^{-zx},
~z,x,u,v \in \C.
\eea
It immediately follows that the $D_{11}^{(N,k)}$ enjoys a determinantal structure, once we know the kernel corresponding to the new weight \eqref{eqnwt}. If the term of unity was not present in the weight, the orthogonal polynomials would follow immediately from a Christoffel type theorem for orthogonal polynomials in the complex plane \cite{AV03}.
Notice that the weight \eqref{eqnwt} is in general complex. Such a situation is not uncommon in non-Hermitian RMT, e.g. when applied to QCD with chemical potential \cite{Misha}, cf. \cite{AOSV} for the orthogonal polynomial approach.  For $k=1$ we have to compute the normalising partition function for this weight, given by the product of their (pseudo) norms. 

Likewise, we consider off-diagonal overlaps conditioned on $k\geq 2$ eigenvalues, compared to $k=2$ in \eqref{offdiagO-def}. Based on \cite{chalker2} for $k=1$  and the permutation symmetry of $p_N(\Lambda)$, we have 
\bea
D_{12}^{(N,k)}(\lambda_1,\ldots,\lambda_k)&=&\frac{N!}{(N-k)!}\int\limits_{\C^{N-k}}\prod_{i=k+1}^{N}
d\lambda_i
d\bar{\lambda}_i\,p_N(\Lambda)\frac{1}{|\lambda_1-\lambda_2|^2}
\nonumber\\
&&\qquad\qquad\qquad\times
\prod_{\ell=3}^N \left[1+\frac{1}{(\lambda_1-\lambda_\ell)
\left(\bar{\lambda}_2-\bar{\lambda}_\ell\right)}\right]
\nonumber\\
&=&\frac{-e^{-|\lambda_1|^2-|\lambda_2|^2}N!}{Z_N(N-k)!}\!\int\limits_{\C^{N-k}}
\!\prod_{i=k+1}^{N}\!d\lambda_i
d\bar{\lambda}_i
\Delta^{(N-1)}(\lambda_2,\ldots, \lambda_N)\nonumber\\
&&\times
\Delta^{(N-1)}(\bar{\lambda}_1,\bar{\lambda}_3,\ldots, \bar{\lambda}_N)
\prod_{m=3}^N\pi \omega(\lambda_m, \bar{\lambda}_m\mid \lambda_1, \bar{\lambda}_2).
\nonumber\\
\label{eqnd12int}
\eea
Also here a determinantal structure arises as a consequence of that for $D_{11}^{(N,k)}$. In the following it will be very important to view all variables $\lambda_j$ and $\bar{\lambda}_j$ 
for $1\leq j\leq k$
as independent. From the  integrals in \eqref{eqnd11int} and \eqref{eqnd12int} containing polynomials and exponentials it is clear that $D_{11}^{(N,k)}$ and $D_{12}^{(N,k)}$ viewed as functions in all their independent $2k$ variables are entire. 

%%%%%%%%%%%%%%%%%%%%%%%%%%%%%%%%%%%%%%%%%%%%%%%%%%  
\section{Results at Finite-$N$}\label{finiteN}

The first observation made in \cite{ATTZ} is a simple operation relating $D_{11}^{(N,k)}$ and $D_{12}^{(N,k)}$, that allows to compute the latter from the former, both at finite and large-$N$. 
Let $\hat{T}$ be the transposition acting on functions $g$ on $\C^{2k}$, with $k \geq 2$, depending on the set of four variables 
$\lambda_1, \bar{\lambda}_1, \lambda_2, \bar{\lambda}_2$ (and possibly more). It is defined by exchanging $\bar{\lambda}_1\leftrightarrow\bar{\lambda}_2$: 
\bea
\hat{T}g(\lambda_1, \bar{\lambda}_1, \lambda_2, \bar{\lambda}_2,\ldots)
=g(\lambda_1, \bar{\lambda}_2, \lambda_2, \bar{\lambda}_1,\ldots).
\eea
In particular it leaves the remaining variables $\lambda_3, \lambdab_3,\ldots \lambda_k,\lambdab_k$  (if present) untouched. This leads to the following relation.

%%%%%%%%%%%%%%%%%%%%%%%%%%%%%%%%%%%%%%%%%%%%%%%%%%%%%%%
\begin{lemma}\cite{ATTZ} Exact relation between conditional diagonal and off-diagonal overlaps.\label{thm_rel}
For any $2\leq k \leq N$, the following identity holds:
\bea\label{rltn}
D_{12}^{(N,k)}(\lambda_1,\ldots,\lambda_k)=\frac{-\ e^{-|\lambda_1-\lambda_2|^2}}
{1-|\lambda_1-\lambda_2|^2}\hat{T}D_{11}^{(N,k)}(\lambda_1,\ldots,\lambda_k).
\eea
\end{lemma}
Notice that in order to determine the off-diagonal overlap of Chalker and Mehlig,  $D_{12}^{(N,2)}(\lambda_1,\lambda_2)$ in \eqref{offdiagO-def}, we need to know the diagonal overlap $D_{11}^{(N,2)}(\lambda_1,\lambda_2)$ conditioned on {\it two} eigenvalues. 
\begin{proof}
Lemma \ref{thm_rel} is easily seen when applying $\hat{T}$ to \eqref{eqnd11int} for $k\geq2$:
\bea
\hat{T}D_{11}^{(N,k)}(\lambda_1,\ldots,\lambda_k)
&=&\frac{N!\ e^{-\lambda_1\bar{\lambda}_2}}{Z_N(N-k)!}\int\limits_{\C^{N-k}}\prod_{i=k+1}^{N}\!
d\lambda_i d\bar{\lambda}_i
\Delta^{(N-1)}(\lambda_2,\ldots, \lambda_N) \nonumber\\
&&\times 
\Delta^{(N-1)}(\bar{\lambda}_1,\bar{\lambda}_3\ldots, \bar{\lambda}_N) 
\ \pi \omega( \lambda_2,\bar{\lambda}_1\mid \lambda_1,\bar{\lambda}_2)
\nonumber\\
&&\times\prod_{m=3}^N \pi \omega( \lambda_m,\bar{\lambda}_m\mid \lambda_1,\bar{\lambda}_2).
\eea
Writing out the first weight that can be pulled out of the integral,
\bea
\pi \omega( \lambda_2,\bar{\lambda}_1\mid \lambda_1,\bar{\lambda}_2)=\left(1+(\lambda_2-\lambda_1)(\bar{\lambda}_1-\bar{\lambda}_2)\right)\ e^{-\lambda_2\bar{\lambda}_1},
\eea
as well as comparing to \eqref{eqnd12int} the statement \eqref{rltn} follows.
\end{proof}

%%%%%%%%%%%%%%%%%%%%%%%%%%%%%%%%%%%%%%%%%%%%%%%%%%
\subsection{Determinantal structure of the conditional diagonal overlaps}\label{detstructure}

Comparing \eqref{cgin1} and \eqref{eqnd11int} and using the theory of Dyson and Mehta \cite{mehta} (which also applies to kernels that are not self adjoint) we can immediately read off the determinantal structure of the $k$-th diagonal overlap:
\bea
\nonumber
D_{11}^{(N,k)} (\lambda_1,\ldots,\lambda_k)&=&\frac{Z^\prime_{N-1}}{Z_N} \,
e^{-|\lambda_1|^2}\det_{2\leq i,j\leq k}
\left[K^{(N-1)}_{11}(\lambda_i,\bar{\lambda}_i,\lambda_j,\bar{\lambda}_j\mid \lambda_1,\bar{\lambda}_1)\right],
\\ \label{eqd11_gen}
\eea
where we have defined the corresponding kernel and reduced kernel 
\bea
K^{(N)}_{11}(x,\bar{x},y,\bar{y}\mid \lambda_1, \bar{\lambda}_1)&=&
\omega(x, \bar{x}\mid \lambda_1, \bar{\lambda}_1)
\ \kappa^{(N)}(\bar{x},y\mid \lambda_1, \bar{\lambda}_1)\label{eq_redk} ,\\
\kappa^{(N)}(\bar{x},y\mid \lambda_1, \bar{\lambda}_1)&=&
\sum_{k=0}^{N-1} \frac{\overline{P_{k}(x)} Q_{k}(y)}{d_k}\ ,
\label{eq_redk2}
\eea
respectively. It contains the monic polynomials $P_k(x)$ and $Q_k(y)$, that are orthogonal  (and in general different) with respect to the weight \eqref{eqnwt}
\bea\label{eq_bop}
\langle P_i, Q_j\rangle :=
\int_{\C} dzd\bar{z}\ \omega(z, \bar{z}
\mid \lambda_1, \bar{\lambda}_1)
\overline{P_i(z)} Q_j(z)=
\delta_{i,j}d_j\ .
\eea
The corresponding partition function follows  in terms of the (squared) norms 
\bea
Z^\prime_{N-1}=(N-1)!\prod_{j=0}^{N-2}d_j\ , 
\label{Z'}
\eea
after applying again Andr\'ei\'ef's integral formula \eqref{Andreief}.  For $k=1$ the determinant in \eqref{eqd11_gen} is absent - a notation we will adopt throughout - and the diagonal overlap  $D_{11}^{(N,1)} (\lambda_1)$ is only determined through these pre-factors in \eqref{eqd11_gen}.

An alternative representation of the reduced kernel uses the inverse $C_{ij}^{(N-1)}$ of the moment matrix 
$\mathcal{M}_{ij}=\langle z^i, z^j\rangle,~0\leq i,j\leq N-1$, 
leading to \cite{borodin}
\bea\label{eq_redker}
\kappa^{(N)} (\bar{z}, z\mid \lambda_1, \bar{\lambda}_1)=\sum_{i,j=0}^{N-1} z^i C^{(N-1)}_{ij} \bar{z}^j . 
\eea
Employing an  $LDU$-decomposition of $\mathcal{M}$, 
\bea\label{eq_ldu}
\mathcal{M}=LDU.
\eea
where $D$ is a diagonal matrix, $L$ and $U^T$ are lower
triangular matrices with the diagonal entries equal to $1$,  we can express the above polynomials and norms in terms of these matrices
\bea\label{eq_bop1}
P_{k}(z)&=&\sum_{m=0}^{k} (\bar{L}^{-1})_{km}z^m,
\\
\nonumber
Q_{k}(z)&=&\sum_{m=0}^{k} z^m(U^{-1})_{mk},
\eea
for $k\geq 0$, with $D=\mbox{diag}(d_0,\ldots,d_{N-1})$.
In \cite{ATTZ} Section 3.4 the matrices $L,D$, and $U$ were determined. They can be expressed in terms of the following function
\bea\label{fpols}
f_{p}(x)=(p+1)e_{p}(x)-x e_{p-1}(x),~p=0,1,\ldots ,
\eea
containing the exponential polynomials 
\begin{eqnarray}\label{expols}
e_{p}(x)=\sum_{k=0}^{p} \frac{x^k}{k!}, ~p=0,1,2,\ldots\ , 
\end{eqnarray}
where we define $e_{-1}(x)\equiv0$. The resulting expressions read:
\bea\label{eq_ldul}
L_{pm}&=&\delta_{pm}-\bar{\lambda}_1\frac{f_{p-1}(\lambda_1 \bar{\lambda}_1)}{f_p(\lambda_1\bar{\lambda}_1)} \delta_{p,m+1},~p,m \geq 0,\\
\label{eq_ldud}
d_{m}&=& (m+1)! \frac{f_{m+1}(\lambda_1 \bar{\lambda}_1)}{f_m(\lambda_1 \bar{\lambda}_1)},~m\geq 0,\\
 U_{mq}&=&\delta_{mq}-\lambda 
 \frac{f_{m-1}(\lambda \bar{\lambda})}{f_m(\lambda\bar{\lambda})} \delta_{q,m+1},~m,q \geq 0\ .\label{eq_lduu}
\eea
This immediately determines the normalisation constant \eqref{Z'} and thus the prefactor in \eqref{eqd11_gen}. In particular it gives an exact finite-$N$ expression for the diagonal overlap at $k=1$ \eqref{diagO-def} 
\bea
D_{11}^{(N,1)} (\lambda_1)&=&\frac{Z^\prime_{N-1}}{Z_N} 
e^{-|\lambda_1|^2} =\frac{1}{\pi} f_{N-1}(\lambda_1\bar{\lambda}_1)\ e^{-\lambda_1\bar{\lambda}_1}\ ,
\eea
after multiplying out the telescopic product of the norms in \eqref{Z'}. The inversion of the lower triangular matrices $L$ and $U^T$ is also not difficult, resulting into
\bea
(L^{-1})_{pq}&=&\left\{
\begin{array}{cc}
0& q>p,\\
\bar{\lambda}_1^{p-q} \frac{f_q(\lambda_1\bar{\lambda}_1)}{f_p(\lambda_1\bar{\lambda}_1)}&q\leq p,
\end{array}
\right.
\label{eq_luinv}\\ 
(U^{-1})_{pq}&=&\left\{
\begin{array}{cc}
\lambda_1^{q-p} \frac{f_p(\lambda_1\bar{\lambda}_1)}{f_q(\lambda_1\bar{\lambda}_1)}& q\geq p,\\
0&q<p.
\end{array}
\right.
\nonumber
\eea 
While this determines the polynomials \eqref{eq_bop1} and thus also the kernel \eqref{eq_redk} this does not lead to a form that is easily amenable to an asymptotic large-$N$ analysis. The reason is that the polynomials \eqref{eq_bop1} are not standard polynomials, with existing tables for their asymptotic behaviour. Thus the  reduced kernel in \eqref{eq_redk2} containing a triple sum is not easy to handle in the limit $N\to\infty$. 

Fortunately, in \cite{ATTZ} an alternative form was derived after a long calculation.
It only contains single sums in terms of the exponential polynomials \eqref{expols} and the function \eqref{fpols}. Defining the function 
\bea\nonumber
\frak{F}_n(x,y,z)&=&e_n(xy)\cdot e_n(xz)-e_n(xyz)\cdot e_n(x)\cdot \left(1-x(1-y)(1-z)\right)\\
&&+\frac{(1-y)(1-z)}{n!}\cdot \frac{(xyz)^{n+1}e_{n}(x)-x^{n+1}e_{n}(xyz)}{1-yz},
\label{total}
\eea
for $~n=0,1,\ldots$, this result can be cast into the following 
%%%%%%%%%%%%%%%%%%%%%%%%%%%%%%%%%%%%%%%%%%%%%%%%%%%
\begin{thm} \cite{ATTZ} Determinantal structure of conditional diagonal overlaps.\label{thm_fn}
For any $1\leq k\leq N$, it holds
\bea
\label{ovlpn11}
D_{11}^{(N,k)}(\lambda_1,\ldots,\lambda_k)&=&\frac{1}{\pi}f_{N-1}(|\lambda_1|^2)
e^{-|\lambda_1|^2}\\
&&\times 
\det_{2\leq i,j\leq k}\left[K^{(N-1)}_{11}\left(\lambda_i,\bar{\lambda}_i,\lambda_j,\bar{\lambda}_j\! \mid\!
\lambda_1,\bar{\lambda}_1\right)\right],
\nonumber
\eea
where the kernel $K^{(N-1)}_{11}$ from \eqref{eq_redk} is given in terms of  the weight \eqref{eqnwt} 
and the reduced kernel 
\bea
\kappa^{(N)}
(\xb,y\!\mid\! \lambda_1, \bar{\lambda}_1)=
\frac{\left(N+1\right) \frak{F}_{N+1}\big(\lambda_1 \bar{\lambda}_1,\frac{\bar{x}}{\overline{\lambda}_1},\frac{y}{\lambda_1}\big)
-\lambda_1 \bar{\lambda}_1\frak{F}_{N}\big(\lambda_1 \bar{\lambda}_1,\frac{\bar{x}}{\overline{\lambda}_1},\frac{y}{\lambda_1}\big)}
{\left(\bar{x}-\overline{\lambda}_1\right)^2\left(y-\lambda_1\right)^2 f_{N}\left(\lambda_1 \overline{\lambda}_1\right)}.
\nonumber\\
\label{thm1_redkern}
\eea
\end{thm}
The result for the conditional off-diagonal overlaps \eqref{eqnd12int} follows from Lemma \ref{thm_rel} and we refer to \cite{ATTZ} for details of its determinantal structure, given in terms of a matrix valued $2\times 2$ kernel that follows from Theorem \ref{thm_fn}.

%%%%%%%%%%%%%%%%%%%%%%%%%%%%%%%%%%%%%%%%%%%%%%%%%%  
\section{Bulk Universality at Large-$N$}\label{bulklargeN}

In this section we focus on a particular large-$N$ limit, the local scaling limit in the  bulk of the spectrum. Compared to \cite{ATTZ} where this limit was only taken at the origin - a point that is representative for the bulk spectrum in the complex Ginibre ensemble - we generalise the result to any bulk point. In \cite{ATTZ} many further results we obtained at large-$N$ based on Theorem \ref{thm_fn}, including the large-argument limit of the local bulk correlations and the local edge scaling limit. We refer to \cite{ATTZ} for the precise statements. Once again we only focus on the diagonal overlap. Fixing a bulk point  $\sqrt{N}z_0$, with $0\leq |z_0|<1$, cf. \eqref{circular},  we define the following bulk scaling limit
\bea
D_{11}^{(bulk,\,k)}(\lambda_1,\ldots,\lambda_k)=\lim_{N\rightarrow \infty} \frac{1}{N} D_{11}^{(N,k)}(\sqrt{N}z_0+\lambda_1,\ldots,\sqrt{N}z_0+\lambda_k)\ ,\quad
\eea
and correspondingly for the off-diagonal overlap $D_{12}^{(bulk,\,k)}$.

%%%%%%%%%%%%%%%%%%%%%%%%%%%%%%%%%%%%%%%%%%%%%%%%%%%
\begin{thm}\label{thm_bulk}
Local bulk scaling limit of conditional diagonal overlaps. It holds that
\bea
D_{11}^{(bulk,\,k)}(\lambda_1,\ldots,\lambda_k)=
\frac{1}{\pi} \det_{2\leq i,j\leq k}
\left[K^{(bulk)}_{11}(\lambda_i,
\bar{\lambda}_i,
\lambda_j,\bar{\lambda}_j\mid \lambda_1,\bar{\lambda}_1)
\right]\label{ovlp11},
\eea
where the limiting kernel is given by 
\bea\label{thm_k11bulk}
K^{(bulk)}_{11}(u,\bar{u},v,\bar{v}\mid \lambda,\bar{\lambda})=
\frac{1}{\pi} \left(1+|u-\lambda|^2\right)
e^{-|u-\lambda|^2}
\kappa^{(bulk)}(\bar{u},v\mid \lambda, \bar{\lambda})\ ,\quad
\eea
together with the limiting reduced kernel
\bea\label{thm_redker11bulk}
\kappa^{(bulk)}(\bar{u},v\mid \lambda, \bar{\lambda})=
\left.
\frac{d}{dz}\left(\frac{e^z-1}{z}\right)
\right|_{z=(\bar{u}-\bar{\lambda})(v-\lambda)}.
\eea
\end{thm}
A similar result can be derived for $D_{12}^{(bulk,\,k)}$ using again Lemma \ref{thm_rel}, and we refer to \cite{ATTZ} for details. 
\begin{proof}
The proof of the theorem for $z_0=0$ can be found in \cite{ATTZ}.
Let $0<z_0<1$. Note that
\bea
e^{-x} e_{N-1}(x) = \frac{\Gamma(N,x)}{\Gamma(N)}\ ,
\label{epolygamma}
\eea
that relates the exponential polynomial to the incomplete Gamma function 
$\Gamma(N,x)=\int_x^\infty dt t^{N-1}e^{-t}$. In particular it holds for large-$N$  \cite{NIST}  that 
\bea
\frac{\Gamma(N,N|z|^2)}{\Gamma(N)}\sim \Theta(1-|z|^2) \ ,
\label{expasympt}
\eea
given in terms of the Heaviside function $\Theta$.
The uniform convergence on the disc of radius $\sqrt{N}$ implies that the exponential polynomial $e_N(x)$ can be replaced by the exponential $e^x$
away from the edge of the spectrum\footnote{For a finer asymptotic at the edge  see \cite{NIST}, leading to the local complementary error function kernel stated in \cite{TaoVu}, including its universality.}. We will simply go through the asymptotic of the building blocks of the prefactor and kernel in \eqref{ovlpn11} and \eqref{thm1_redkern}, keeping the leading order terms. We denote by 
\bea
\lambda=\lambda_1&=& \sqrt{N}z_0+\rho \ ,\nonumber\\
x&=& \sqrt{N}z_0+\xi \ ,\nonumber\\
y&=& \sqrt{N}z_0+\eta \ .
\label{scalingvar}
\eea
 For the exponential polynomial we have 
\bea
e_{N\pm1}(|\lambda|^2)\sim e^{|\lambda|^2}= \exp\left[N|z_0|^2+ \sqrt{N}(z_0\bar{\rho}+\bar{z}_0\rho) +|\rho|^2\right],
\label{eNasympt}
\eea
which implies for \eqref{fpols}
\bea
\frac{1}{N}f_{N-1}(|\lambda|^2)\sim \big(1-|z_0|^2\big)
\exp\left[N|z_0|^2+ \sqrt{N}(z_0\bar{\rho}+\bar{z}_0\rho) +|\rho|^2\right]
.
\label{fasympt}
\eea
The following term requires a bit more analysis:
\bea
\label{Fasympt}
\frak{F}_{N+1}\left(|\lambda|^2,\frac{\bar{x}}{\bar{\lambda}},\frac{y}{\lambda}\right)&\sim&e^{\lambda\bar{x}+\bar{\lambda}y}-e^{\bar{x}y+|\lambda|^2}\big(1-(\bar{\lambda}-\bar{x})(\lambda-y)\big)\\
&&+\frac{(\bar{\lambda}-\bar{x})(\lambda-y)}{(|\lambda|^2-\bar{x}y)(N+1)!}
\left( (\bar{x}y)^{N+2} e^{|\lambda|^2}- (\lambda\bar{\lambda})^{N+2}e^{\bar{x}y}\right).
\nonumber
\eea
The leading order exponent of the terms in the first line is $2N|z_0|^2$, compared to the leading exponent of the second line $N+N|z_0|^2+(N+2)\ln|z_0|^2$, after using Stirling's formula. The 
latter is thus subleading for large-$N$, due to $t-1>\ln(t)$ for $0<t=|z_0|^2<1$, and can thus be neglected compared to the first.  
The limit of a vanishing denominator (and numerator) can be estimated using 
l'H\^opital, leading to the same conclusion. 
The same argument applies to $|\lambda|^2\frak{F}_{N}$. Together with the asymptotic \eqref{fasympt} this leads to the following asymptotic for the reduced kernel 
\bea
\kappa_{N}(\bar{x},y|\lambda,\bar{\lambda})&\sim& 
e^{N|z_0|^2+\sqrt{N}(z_0\bar{\xi}+\bar{z_0}\eta)-|\rho|^2}
\nonumber\\
&&\times
\frac{e^{\rho\bar{\xi}+\bar{\rho}\eta}- e^{\bar{\xi}\eta+|\rho|^2}\big(1- (\bar{\xi}-\bar{\rho})(\eta-\rho)\big)}{(\bar{\xi}-\bar{\rho})^2(\eta-\rho)^2}.
\nonumber\\
\eea
The weight function \eqref{eqnwt} in the scaling \eqref{scalingvar} reads
\bea
w(x,\bar{x}|\lambda,\bar{\lambda})= \frac{1}{\pi}\big(1+(\xi-\rho)(\bar{\xi}-\bar{\rho}) \big)e^{-N|z_0|^2-\sqrt{N}(z_0\bar{\xi}+\bar{z_0}\xi)-|\xi|^2}.
\label{weightlim}
\eea
Putting all these factors together we obtain
\bea
K_{11}^{(N)}(x,\bar{x},y,\bar{y}|\lambda,\bar{\lambda})&\sim&
e^{\sqrt{N}\bar{z}_0(\eta-\xi) }
\frac{e^{-|\xi|^2-|\rho|^2}
\big(1+|\xi-\rho|^2 \big)}{\pi(\bar{\xi}-\bar{\rho})^2(\eta-\rho)^2}
\nonumber\\
&&\times\left( e^{\rho\bar{\xi}+ \bar{\rho}\eta}- e^{\bar{\xi}\eta+|\rho|^2}
\big(1- (\bar{\rho}-\bar{\xi})(\rho-\eta)\big)
\right) \nonumber \\
&=&
e^{(\sqrt{N}\bar{z}_0+\bar{\rho})(\eta-\xi) }e^{-|\xi-\rho|^2}
\frac{
\big(1+|\xi-\rho|^2 \big)}{\pi}
\nonumber\\
&&\times\frac{\left( 1- e^{(\bar{\rho}-\bar{\xi})(\rho-\eta)}
\big(1- (\bar{\rho}-\bar{\xi})(\rho-\eta)\big)
\right)}{(\bar{\xi}-\bar{\rho})^2(\eta-\rho)^2}.
\label{K11asympt}
\eea
The prefactors $f_N(\eta)= \exp[(\sqrt{N}\bar{z}_0+\bar{\rho})\eta]$ and $1/f_N(\xi)$ can be
eliminated by the conjugation of the kernel  \eqref{K11asympt} and thus the kernel in \eqref{K11asympt} is equivalent to the kernel \eqref{ovlp11} stated in our Theorem \ref{thm_bulk}. 
\end{proof}
We note in passing  that from \eqref{cgin1},  
\bea\label{circular}
\rho^{N,1}(\sqrt{N}z)=K_{ev}^{(N)}(\sqrt{N}z,\sqrt{N}z)=\frac{1}{\pi} \frac{\Gamma(N,N|z|^2)}{\Gamma(N)}=\frac1\pi \Theta(1-|z_0|^2)\ , \qquad
\eea
the asymptotic \eqref{expasympt} leads to 
the circular law for the global density. 

In \cite{ATTZ} several further results were derived, including the local edge scaling limit of the conditional diagonal and off-diagonal overlap, the asymptotic connection between these local edge and bulk correlations, as well as the algebraic decay of the local bulk correlations in the large-argument limit. We shall not repeat these results here. We only mention that the following relation was derived for the large-argument limit in \cite{ATTZ}, relating the local bulk $k$-point density and overlap correlation functions. 

\begin{lemma}\cite{ATTZ} Relation between conditional diagonal overlap and density correlations in the local bulk scaling limit.\label{La-rho-D11}
For $k\geq2$ it holds that
\bea
\label{prodd11}
&&D_{11}^{(bulk,\,k)}(\lambda_1,\ldots,\lambda_k)=(-1)^{k-1} \prod_{m=2}^k \frac{1+|\lambda_m-\lambda_{1}|^2}{|\lambda_m-\lambda_{1}|^4}
\\
&&\qquad\qquad\times 
\left(1-|\lambda_m-\lambda_{1}|^2-(\lambda_m-\lambda_{1})\frac{\partial}{\partial \lambda_{m}}\right)\rho^{(bulk,\, k)}(\lambda_1,\ldots,\lambda_k). 
\nonumber
\eea
\end{lemma}
For $k=1$ the overlap is constant $D_{11}^{(bulk,\,1)}(\lambda_1)=\frac{1}{\pi}$, cf. Theorem \ref{thm_bulk}, and equals the local (and global) bulk density $\rho^{(bulk,\,1)}$, cf. \eqref{rhokbulk} below. 
The limiting $k$-point density correlation functions in \eqref{prodd11} are summarised in the following
\begin{thm} Local limiting bulk correlation functions. 
\label{localbulkrho}
For any  bulk point 
$\sqrt{N}z_0$, with $0\leq |z_0|<1$, the following bulk scaling limit holds
\bea
\rho^{(bulk,\,k)}(\lambda_1,\ldots,\lambda_k)&=&\lim_{N\rightarrow \infty} \rho^{(N,k)}(\sqrt{N}z_0+\lambda_1,\ldots,\sqrt{N}z_0+\lambda_k)
\nonumber\\
&=& \det_{1\leq i,j\leq k}\left[K_{Gin}(\lambda_i,\lambda_j)\right],
\label{rhokbulk}
\eea
with the Ginibre kernel 
\bea
K_{Gin}(x,y)=\frac{1}{\pi}\exp\left[-\frac12|x|^2-\frac12 |y|^2+\bar{x}y\right].
\eea
\end{thm}
A similar result to Lemma \ref{La-rho-D11} holds for $D_{12}^{(bulk,\,k)}$, thanks to Lemma \ref{thm_rel}.
For the proof of Lemma \ref{La-rho-D11} we refer to \cite{ATTZ}. The first version of Theorem \ref{localbulkrho} goes back to \cite{ginibre} who proved this limit at the origin. The extension to the bulk is not difficult, using \eqref{epolygamma} und \eqref{expasympt}.

%%%%%%%%%%%%%%%%%%%%%%%%%%%%%%%%%%%%%%%%%%%%%%%%%%  
\section{Discussion and Open Problems}\label{disc}

In these proceedings we have reported on recent results regarding the determinantal structure for the conditional overlaps in the complex Ginibre ensemble \cite{ATTZ}. 
The analyticity of the diagonal overlap yields the off-diagonal overlap using a simple transposition operator in Lemma \ref{thm_rel}. In the large-$N$ limit we focussed on the local bulk scaling limit, generalising the results of \cite{ATTZ} to hold for arbitrary bulk points, to which the overlaps are conditioned. We expect that this universality holds for a more general class of weight functions. Given that the local bulk universality of the $k$-point density correlation functions Theorem \ref{localbulkrho} is known to hold for a much wider class of random matrices, see e.g. \cite{TaoVu} for Wigner matrices, and that we have the relation that relates the corresponding overlap to the density correlation functions through a differential operator in Lemma \ref{La-rho-D11} in the Gaussian case, we conjecture the local bulk overlaps to be universal as well. The same probably holds for the local overlap edge correlations, cf. \cite{GW}, even if we currently do not have such a relation to the density correlations. The latter are universal as well, cf. \cite{TaoVu}.

It is not uncommon that an approach that uses orthogonal polynomials for finite-$N$ is more difficult when it comes to deriving global correlation functions, both for eigenvalues and eigenvectors. Here, the methods of Green's functions \cite{BurdaPRL,Burda+} and Feynman diagrams \cite{MAN-WT} advocated by the Krakow group have proven much more useful. We refer to \cite{MAN-WT} for a comprehensive list of global bulk correlations of the off-diagonal overlap in various ensembles, see also the references therein. 
It seems clear that the global bulk correlations are much less universal, containing a weight specific term, and that only their algebraic decay that also follows from the  local bulk correlations at large argument \cite[Corollary 4]{ATTZ} is universal. Clearly much more future study is needed on this aspect, including more general weight functions. A particularly interesting ensemble is the elliptic Ginibre ensemble that allows one to interpolate between the Ginibre and Gaussian unitary ensemble. It would be very interesting to study the transition from the strongly correlated overlaps between eigenvectors and their known independence in the Hermitian limit.  For the real eigenvalues of the real Ginibre ensemble results for the elliptic ensemble have been reported very recently \cite{YVF-WT} - see also these proceedings.

Let us comment further on the relation to known results in the other two Ginibre ensembles with real and quaternionic entries. Away from the real axis, the complex eigenvalue correlation functions at the edge and in the bulk of the spectrum have been shown to agree between the real, complex and quaternionic Ginibre ensemble, see \cite{Rider,BorodinSinclair} and \cite{BorodinSinclair,AKMP}, respectively. Regarding eigenvectors their global bulk correlations have been shown to agree for the complex and quanternionic ensemble \cite{AFK}. Therefore, it is tempting to conjecture that the agreements holds also for the real ensemble, away from the real line. Here and in the quaternionic ensemble the local overlap correlations in the bulk and at the edge are currently open. It would be very helpful to detect an integrable structure for the real and quaternionic ensemble  at finite-$N$, this time of Pfaffian type, as reported in a determinantal form for the complex  ensemble here. At least in the quaternionic Ginibre ensemble the way to proceed using skew-orthogonal polynomials is clear, based on the expression for the overlaps in terms of averages over complex eigenvalue pairs only, see \cite{AFK}. 

Research on eigenvector statistics has become a very active field now and we hope that this paper will lead to fruitful applications in the area of the theme of this workshop.

\section*{Acknowledgements}
Support from the following grants is gratefully acknowledged:
The Knut and Alice Wallenberg Foundation and 
CRC1283 ``Taming uncertainty and profiting from randomness and low regularity in analysis, stochastics and their applications"  by  
 the German Research Foundation DFG for Gernot Akemann, Roger Tribe is partially supported by a Leverhulme Research Fellowship RF-2-16-655.

%%%%%%%%%%%%%%%%%%%%%%%%%%%%%%%%%%%%%%%%%%%%

\end{document}